\documentclass[12pt]{article}
\newcommand{\ii}{\'{\i}}
\newcommand{\bb}{$\beta\beta$}
\newcommand{\bt}{$\beta\beta_{2\nu}$}
\newcommand{\bz}{$\beta\beta_{0\nu}$}
\newcommand{\gd}{$^{160}$Gd}

\begin{document}

\title{Selection rules in the \bb ~decay of deformed nuclei}
\author{Jorge G. Hirsch \thanks{hirsch@nuclecu.unam.mx},
Octavio Casta\~nos \thanks{ocasta@nuclecu.unam.mx},
Peter O. Hess \thanks{hess@nuclecu.unam.mx}\\
{\small\it Instituto de Ciencias Nucleares, Universidad Nacional Aut\'onoma de
M\'exico,}\\ {\small\it A. P. 70-543 M\'exico 04510 D.F.}
\and
Osvaldo Civitarese \thanks{civitare@venus.fisica.unlp.edu.ar}
\\
{\small\it Departamento de F\ii sica, Universidad Nacional de La Plata,}
\\
{\small\it  c.c.67; 1900, La Plata, Argentina}
  }

\maketitle

\begin{abstract}
The $2\nu$ \bb ~decay half-lives of six nuclei, whose decays were
previously reported as theoretically forbidden, are calculated by
including the pairing interaction, which mixes different
occupations and opens up the possibility of the decay. All allowed
channels for the $0\nu$ \bb ~decay are also computed. The
estimated $2 \nu$ \bb ~half-lives suggest that measurements in
$^{244}$Pu may find positive signals, and that planned experiments
would succeed in detecting the \bt ~decay in \gd. Limits for the
zero neutrino mode, in the analyzed deformed emitters, are
predicted.

\noindent
Pacs numbers: 21.60.Fw, 23.40.Hc, 27.70.+q, 27.90.+b
\end{abstract}

Neutrinoless double beta decay (\bz), if detected,
would offer definitive evidence that the neutrino is a Majorana particle,
i.e. that it is its own antiparticle \cite{Ver86,Suh98}.
It would also provide the information needed to determine neutrino masses,
complementary to the one obtained from solar and atmospheric neutrino
experiments \cite{SK01,SNO01}.

Theoretical nuclear matrix elements are needed to convert
experimental half-life limits, which are available for many
$\beta\beta$-unstable isotopes \cite{Suh98,Deb01}, into constrains
for the effective Majorana mass of the neutrino and the
contribution of right-handed currents to the weak interactions.
Thus, these matrix elements are essential to understand the
underlying physics \cite{Vog86,Civ87,Rad96}.
The two neutrino mode of the double beta decay (\bt ) is allowed
as a second order process in the standard model. It has been
detected in ten nuclei \cite{Deb01} and it has served to test a
variety of nuclear models \cite{Suh98}.

The pseudo SU(3) approach has been used to describe many low-lying
rotational bands, as well as B(E2) and B(M1) intensities, in rare
earth and actinide nuclei, both with even-even and odd-mass
numbers. The theoretical results show, in general, a very good
agreement with the data \cite{Var00a}. The $\beta\beta$ half-lives
of some heavy deformed nuclei, which may decay to the ground and
excited states of the daughters, were evaluated for the two and
zero neutrino emitting modes
\cite{Cas94,Hir95a,Hir95b,Hir95c,Hir95d} using the pseudo SU(3)
scheme. The predictions were in good agreement with the available
experimental data for $^{150}$Nd and $^{238}$U. The double
electron capture decay channel was studied for the decay of other
three nuclei \cite{Cer99}.

The simplest pseudo SU(3) model predicts the complete suppression
of the \bt ~decay for the following five nuclei: $^{154}$Sm,
$^{160}$Gd, $^{176}$Yb, $^{232}$Th and $^{244}$Pu
\cite{Cas94,Hir95d}. Recently, it was argued that the cancellation
of the \bt ~decay in $^{160}$Gd would suppress the background for
the detection of the 0$\nu$ mode \cite{Dan00}.

In the present contribution we extend the previous research
\cite{Cas94,Hir95a,Hir95b,Hir95c,Hir95d,Hir01} and  evaluate the
$\beta\beta$ half-lives of $^{154}$Sm, $^{160}$Gd, $^{170}$Er,
$^{176}$Yb, $^{232}$Th and $^{244}$Pu using the pseudo SU(3)
model. In these nuclei the 2$\nu \beta \beta$ mode is forbidden
when the most probable occupations are considered. To be able to
evaluate finite half-lives, we were forced to include in the
calculations states with different occupation numbers which can be
mixed through the pairing interaction. The amount of this mixing
is evaluated, and the possibility of observing the $\beta\beta$
decay is discussed for both the 2$\nu$ and 0$\nu$ modes.

Pseudo-spin symmetry \cite{Blo95} describes the quasi-degeneracy
of the single-particle orbitals with $j = l - 1/2$ and $j = (l-2)
+ 1/2$ in the $\eta$ shell. It allows these orbitals to be
classified as pseudo-spin partners with quantum numbers $\tilde j
= j$, $\tilde\eta = \eta -1$ and $\tilde l = l - 1$.
The first step in the pseudo SU(3) description of any nuclei is to
find the occupation numbers for protons ($p$) and neutrons ($n$)
in the normal and abnormal parity states $n^N_p, n^N_n, n^A_p,
n^A_n$. These numbers are determined by filling the Nilsson levels
from below, as discussed in \cite{Cas94}. The deformations
\cite{Moll95} and occupancies for the 12 isotopes studied in the
present work are shown in Table 1.

\begin{table}
\begin{center}
\begin{tabular}{lc|cccc|ccc}
 Nucleus    &~~$\epsilon_2$ ~~ &~$n_p^N$&$n_p^A$&$n_n^N$&$n_n^A$
& $\Delta E$ & ~$h_{pair}$ & ~~$x_i$~~ \\ \hline
$^{154}$Sm (1)    &   0.250    & 8   & 4    &   6   &   4   &   &      &  0.933\\
$^{154}$Sm (2) &            & 8   & 4    &   8   &   2   & 1.27  &  0.575 &  0.359\\
$^{154}$Gd     &   0.225    & 8   & 6    &   6   &   2   &    &    &  1.000\\
\hline
$^{160}$Gd     &   0.258    & 8   & 6    &   8   &   6   &   &      & 1.000\\
$^{160}$Dy (1) &   0.250    & 10  & 6    &   6   &   6   &    &    &  0.923\\
$^{160}$Dy (2) &            & 8   & 8    &   6   &   6   & 1.71  &  0.865 &  0.385\\
\hline
$^{170}$Er (1) &   0.267    & 10  & 8    &  12   &   8   &   &      & 0.934\\
$^{170}$Er (2) &            & 12  & 6    &  12   &   8   & 1.24  &  0.554 &  0.356\\
$^{170}$Yb     &   0.267    & 12  & 8    &  10   &   8   &    &    &  1.000\\
\hline
$^{176}$Yb     &   0.250    & 12  & 8    &  16   &   8   &   &      & 1.000\\
$^{176}$Hf (1) &   0.250    & 14  & 8    &  14   &   8   &    &    &  0.943\\
$^{176}$Hf (2) &            & 12  & 10   &  14   &   8   & 1.23  &  0.493 &  0.332\\
\hline
$^{232}$Th (1) &   0.192    & 4   & 4    &  10   &   6   &   &      &  0.829\\
$^{232}$Th (2) &            & 6   & 2    &  10   &   6   & 0.318  &  0.391 &  0.559\\
$^{232}$U      &   0.192    & 6   & 4    &   8   &   6   &    &    &  1.000\\
\hline
$^{244}$Pu (1) &   0.208    & 6   & 6    &  16   &   8   &   &      &  0.740\\
$^{244}$Pu (2) &            & 8   & 4    &  16   &   8   & 0.080  &  0.419 &  0.673\\
$^{244}$Cm     &   0.217    & 8   & 6    &  14   &   8   &    &    &  1.000\\
\end{tabular}
\label{occup}
\end{center}
\caption{Deformations, proton and neutron occupation numbers, pairing mixing
$h_{pair}$ and excitation $\Delta E$ energies (both in MeV) and mixing coefficients
$x_i$.}
\end{table}

In the \bt ~decay each Gamow-Teller operator annihilates a proton
and creates a neutron in the same oscillator shell and with the
same orbital angular momentum. As a consequence, the \bt ~decay is
allowed only if the occupation numbers fulfill the following
relationships
\begin{eqnarray}
n^A_{p ,f} = n^A_{p ,i} + 2 ~, &n^A_{n ,f} = n^A_{n ,i} ~, \nonumber \\
n^N_{p ,f} = n^N_{p ,i} ~,  &n^N_{n ,f} = n^N_{n ,i} - 2
~
.\label{num}
\end{eqnarray}

This selection rule \cite{Cas94} forbids the \bt ~decay between the nuclei
marked with (1) or without comments in Table 1.
However, the pairing interaction allows the mixing between states
in the same nuclei with pairs of nucleons transferred between different configurations.
These excited configurations are indicated by (2) in Table 1.
An energy difference $\Delta E$ is required to promote a pair of nucleons from the
last occupied normal parity orbital to the next intruder orbital (or viceversa),
in the deformed single particle Nilsson scheme. It is listed in the seventh column
of Table 1.

The ``leading SU(3) irreps'', those which are the most bounded under the quadrupole -
quadrupole interaction, are the dominant component of the ground state wave function
in these heavy deformed nuclei, representing  in most cases more than 60\% of the
total wave function \cite{Var00a}.
For the eight rare earth isotopes listed in Table 1,
this dominant wave function component can be written as
\begin{eqnarray}
|\hbox{Nucleus}, 0^+ \rangle  =
 | \ (h_{11/2})^{n^A_p},  \ J^A_p =  0; \ (i_{13/2})^{n^A_n} ,\ J^A_n =  0
\rangle_A \label{wf}\\
 | \{2^{\frac {n^N_p}{2}}\} (\lambda_p ,\mu_p); \{2^{\frac {n^N_n}{2}}\}
(\lambda_n ,\mu_n); \ 1 (\lambda ,\mu ) K=1, J = 0 \rangle_N  .
\nonumber
\end{eqnarray}
For the actinide isotopes, the intruder sector is $| \
(i_{13/2})^{n^A_p},  \ J^A_p =  0; \ (j_{15/2})^{n^A_n} ,\ J^A_n =
0 \rangle$.

The proton, neutron, and total SU(3) irreps associated to each set of occupation numbers
are listed in Table 2.
\begin{table}
\begin{center}
\begin{tabular}{l|ccc}
 Nucleus    &$(\lambda_p ,\mu_p)$  &$(\lambda_n ,\mu_n)$  &$(\lambda ,\mu )$
 \\ \hline
$^{154}$Sm (1) &   (10, 4) & (18, 0)   &  (28, 4)\\
$^{154}$Sm (2) &   (10, 4) & (18, 4)   &  (28, 8)\\
$^{154}$Gd     &   (10, 4) & (18, 0)   &  (28, 4)\\
\hline
$^{160}$Gd     &   (10, 4) & (18, 4)   &  (28, 8)\\
$^{160}$Dy (1) &   (10, 4) & (18, 0)   &  (28, 4)\\
$^{160}$Dy (2) &   (10, 4) & (18, 0)   &  (28, 4)\\
\hline
$^{170}$Er (1) &    (10,4) & (24, 0)   &  (34, 4)\\
$^{170}$Er (2) &    (4,10) & (24, 0)   &  (28, 10)\\
$^{170}$Yb     &    (4,10) & (20, 4)   &  (24, 14)\\
\hline
$^{176}$Yb     &    (4, 10) & (18, 8)   &  (22, 18)\\
$^{176}$Hf (1) &    (0, 12) & (20, 6)   &  (20, 18)\\
$^{176}$Hf (2) &    (4, 10) & (20, 6)   &  (24, 16)\\
\hline
$^{232}$Th (1) &   (12, 2) & (30, 4)   &  (42, 6)\\
$^{232}$Th (2) &   (18, 0) & (30, 4)   &  (48, 4)\\
$^{232}$U      &   (18, 0) & (26, 4)   &  (44, 4)\\
\hline
$^{244}$Pu (1) &   (18, 0) & (34, 8)   &  (52, 8) \\
$^{244}$Pu (2) &   (18, 4) & (34, 8)   &  (52, 12)\\
$^{244}$Cm     &   (18, 4) & (34, 6)   &  (52, 10)\\
\end{tabular}
\end{center}
\label{irreps}
\caption{The proton, neutron, and total irreps assigned to each nucleus}
\end{table}

As a first approximation, we will describe the ground state of each nucleus
as a linear combination of these two states:
$$
|\hbox{Nucleus}, 0^+ \rangle =  x_1 \,|\hbox{Nucleus} (1), 0^+
\rangle  + x_2 \,|\hbox{Nucleus} (2), 0^+ \rangle ,  \nonumber
$$
with $x_1^2 + x_2^2 = 1$.

Many multipole-multipole pairing type interactions can remove a pair
of nucleons from an unique parity orbital and create another pair in
a normal parity one. In the present approach we are restricting the
pairs of nucleons in intruder orbits to be coupled to J=0, i.e. to
have seniority zero. Under this approximation
the only term in the Hamiltonian which can connect
states with different occupation numbers
in the normal and unique parity
sectors is pairing. In the present case, the
Hamiltonian
matrix has the simple form
\begin{equation}
H = \left(
\begin{array}{cc}  0  & h_{pair}  \\ h_{pair}  &\Delta E
\end{array}
 \right)
~,
\end{equation}
with $ h_{pair} = \langle \hbox{Nucleus} (2), 0^+ | H_{pair}
|\hbox{Nucleus} (1), 0^+ \rangle , $ whose explicit expression is
given in \cite{Hir01}. The values of $h_{pair}, x_1$ and $x_2$ are
presented in the last two columns of Table 1. In the case of
$^{244}$Pu we are using a small deformation \cite{Moll95}, for
which the two configurations listed in Table 1 are nearly
degenerate, and have maximal mixing. It has important consequences
upon its predicted \bb ~half-life.

The inverse half-life of the two
neutrino mode of the $\beta\beta$-decay, \bt ,
can be evaluated as \cite{Doi85}
\begin{equation}
    \left[\tau^{1/2}_{2\nu}(0^+ \rightarrow 0^+)\right]^{-1} =
      G_{2\nu} \ | \ M_{2\nu} \ |^2 \ \ ,
\end{equation}
where $G_{2\nu}$ is a kinematical factor which
depends on $Q_{\beta\beta}$, the total kinetic energy available
in the decay.

The
nuclear matrix element is
\begin{equation}
M_{2\nu} \approx M_{2\nu}^{GT} = \sum_{N}
{ { \langle  0^+_f \ || \, \Gamma \, || \ 1^+_N \rangle \  \langle 1^+_N \
 || \,\Gamma \, ||\  0^+_i \rangle \,} \over{E_f + E_N -E_i} } ~,\label{m1}
\end{equation}
being $ \Gamma$ the Gamow-Teller operator.
 The  energy denominator contains the intermediate $E_N$, initial $E_i$ and
final $E_f$ energies. The kets
$|1^+_N\rangle$  denote intermediate states.

The mathematical expressions needed to evaluate the nuclear matrix
elements of the allowed $g.s. \rightarrow g.s.$ \bb ~decay
in the pseudo SU(3) model were developed
in \cite{Cas94}.
Using the summation method described in \cite{Cas94},
exploiting the fact that the two body terms of the  $\tilde{SU(3)}$
Hamiltonian commutes with the
Gamow-Teller operator \cite{Hir95a},
resuming the infinite series and recoupling the
Gamow-Teller operators, the following expression was found:
\begin{equation}
M_{2\nu}^{GT} = {\sigma(p,n)^2 \over {\cal E} } \langle 0^+_f |
\left[ [a^{\dagger}_p \otimes \tilde a_n ]^1 \otimes
[a^{\dagger}_{p} \otimes \tilde a_{n}]^1 \right]^{J=0} | 0^+_i
\rangle ,\label{m3}
\end{equation}
where $\sigma(p,n)$ are the 1-body Gamow-Teller matrix elements
and the energy denominator ${\cal E}$ is determined by demanding
that the excitation energy of the Isobaric Analog State equals the
difference in Coulomb energies \cite{Cas94,Hir01}.
The index $p,n$ refer to the orbitals $i_{11/2}^p, i_{13/2}^n$ for
$^{154}$Sm, $^{160}$Gd, $^{170}$Er, and $^{176}$Yb, and to the
orbitals  $j_{13/2}^p, j_{15/2}^n$ for $^{232}$Th and $^{244}$Pu.

As it was discussed in \cite{Cas94} Eq. (\ref{m3}) has no free
parameters, being the denominator ${\cal E}$ a well defined
quantity. The reduction to only one term comes as a consequence of
the restricted proton and neutron spaces of the model. The initial
and final ground states are strongly correlated with  a very rich
structure in terms of their shell model components. The evaluation
of the  matrix elements in the normal space of Eq. (\ref{m3}) is
performed by using $SU(3)$ Racah calculus to decouple the proton
and neutron normal irreps, and expanding the annihilation
operators in their SU(3) tensorial components.

For the six potential \bb ~emitters listed in Tables I and II, the
\bt ~decay  can only proceed through the second component of the
ground state wave function, and for this reason it is proportional
to the amplitude $x_2$. Its explicit expression is given in
\cite{Hir01}.

For massive Majorana neutrinos one can perform the integration
over the four-momentum of the exchanged particle and obtain a
neutrino potential, which for a light neutrino ($m_\nu < 10$ MeV)
has the form

\begin{equation}
H(r,\overline E) = {\frac {2R}{p r}} \int_0^\infty dq {\frac
{sin(qr)} {q+\overline E} }  ~,
\end{equation}

\noindent
where $\overline E$ is the average excitation energy of
the intermediate odd-odd nucleus and the nuclear radius $R$ has been
added to make the neutrino potential dimensionless. In the zero neutrino
case this closure approximation is well justified \cite{Pan92}.
The final formula, restricted to the term proportional to the neutrino
mass, is \cite{Tom91,Doi85}
\begin{equation}
(\tau^{1/2}_{0\nu})^{-1} = \left ( {\frac {\langle m_\nu \rangle}{m_e}} \right )^2
G_{0\nu}  M_{0\nu}^2   ~,
\end{equation}

\noindent
where $G_{0\nu}$ is the phase space integral associated with the emission
of the two electrons. The nuclear matrix elements
$M_{0\nu}$ are \cite{Doi85}
\begin{equation}
M_{0\nu} \equiv | M_{0\nu}^{GT} - {\frac{g_V^2}{g_A^2}} M_{0\nu}^{F} |,
\hbox{~with~}
M_{0\nu}^\alpha =  \langle  0^+_f \| O^\alpha \| 0^+_i\rangle,
\end{equation}
\noindent where the kets
$ \vert 0^+_i \rangle$  and $|0^+_f\rangle$ denote
the corresponding initial and final nuclear  states, the quantities $g_V$
and $g_A$ are the dimensionless coupling constants of the vector and
axial vector nuclear currents, and

\begin{eqnarray}
O^{GT} = \sum\limits_{m,n}
\vec\sigma_m
t^-_m \cdot \vec\sigma_n t^-_n H(|\vec r_m - \vec r_n |,\overline E) ~,\\
O^{F} = \sum\limits_{m,n}  t^-_m t^-_n H(|\vec r_m - \vec r_n |,\overline E)  ~,
\nonumber
\end{eqnarray}
\noindent being $\vec\sigma$ the Pauli matrices related with the spin
operator and $t^-$ the isospin lowering operator, which satisfies
$t^-|n\rangle = |p\rangle$. The superscript GT denotes the Gamow-Teller spin-isospin
transfer channel, while F indicates the Fermi isospin one.
In the present work we use the effective value
$({\frac {g_A} {g_V}})^2 = 1.0$ \cite{Vog86}.

Transforming this operator to the pseudo $SU(3)$ space, we arrive to the expression
\begin{equation}
O^\alpha =  O^\alpha_{N_p N_n} +O^\alpha_{N_p A_n} + O^\alpha_{A_p
N_n} +O^\alpha_{A_p A_n}
~,
\end{equation}
\noindent
where the subscript index $NN, \ NA, \dots$ are indicating the
normal or abnormal spaces of the fermion creation and annihilation operators,
respectively.

In previous works \cite{Cas94,Hir95a} we restricted our analysis
to six potential double beta emitters which, within the
approximations of the simplest pseudo SU(3) scheme, were also
decaying via the $2\nu$ mode. They included the observed
$^{150}$Nd $ \rightarrow$ $ ^{150}$Sm and $^{238}$U $\rightarrow$
$ ^{238}$Pu decays. In these cases two neutrons belonging to a
normal parity orbital decay in two protons belonging to an
abnormal parity one. The transition is mediated by the operator
$O^\alpha_{A_p N_n}$.

\begin{table}
\begin{center}
\begin{tabular}{l|ccccc}
 Nuclei  &$Q_{\beta\beta}$[MeV]  &$G_{2\nu}$ [MeV$^2$ yr$^{-1}$]
&$M_{2\nu}^{GT}$ [MeV$^{-1}$]&  $\tau^{1/2}_{2\nu}$ [yr]  \\ \hline
$^{154}$Sm  & 1.251 &  4.872 10$^{-21}$
& 0.0445  & 1.04 10$^{23}$  \\
$^{160}$Gd  & 1.730 &  8.028 10$^{-20}$
& 0.0455  & 6.02 10$^{21}$  \\
$^{170}$Er  & 0.654 &  6.496 10$^{-23}$
& 0.0374  & 1.10 10$^{25}$\\
$^{176}$Yb  & 1.086 &  3.866 10$^{-21}$
& 0.0306  & 2.77 10$^{23}$\\
$^{232}$Th  & 0.858 &  7.410 10$^{-21}$
& 0.0504  & 5.30 10$^{22}$ \\
$^{244}$Pu  & 1.352 &  4.081 10$^{-19}$
& 0.0617  & 6.43 10$^{20}$ \\
\end{tabular}
\end{center}
\label{2nu}
\caption{The Q-values, phase-space integrals, matrix elements and
predicted half-lives for the \bt ~beta decay}
\end{table}
In Table 3 the \bt ~decays of the six nuclei, previously
reported as forbidden, are presented.
Those with the larger $Q_{\beta\beta}$ values have the larger
phase-space integrals $G_{2\nu}$. The \bt -decay matrix elements
$M_{2\nu}^{GT}$ are suppressed by a factor $x_2 \approx 1/3$
compared with the ``allowed'' ones (see \cite{Cas94}), which
reflects in \bt ~half-lives a factor 10 larger than in other
nuclei with similar Q-values. The exception is $^{244}$Pu, which
for the deformation used has a large mixing, and the shorter \bt
~half-life, which is not far from the limits reported in the
Livermore experiment \cite{Moo92}. The decay of $^{160}$Gd is
suppressed but it is still not far from the present limits
\cite{Bur95}, and large enough to be seen in the proposed
experiments \cite{Dan00}.

Those configurations in which the \bt ~ transitions are forbidden
can still be connected through the zero neutrino mode, due to
presence of the neutrino potential. In this way there are two
terms in the \bz ~decay: one connecting to the basis state which
has allowed \bt ~decay, and one to the state with forbidden \bt
~decay. The equations needed in the first case are the same
employed in the study of allowed decays \cite{Cas94,Hir95a}. The
second one involves the annihilation of two neutrons in normal
parity orbitals, and the creation of two protons in normal parity
orbitals (intruder-intruder in $^{154}$Sm). The transition is
mediated by the operator $O^\alpha_{N_p N_n}$ ($O^\alpha_{A_p
A_n}$). A detailed description of the calculations involved is
presented in \cite{Hir01}. \bz ~phase-space integrals, nuclear
matrix elements and half-lives are shown in Table 4.

\begin{table}
\begin{center}
\begin{tabular}{l|ccc}
 Nucleus  &$G_{0\nu}$ [yr$^{-1}$] &$M_{0\nu}$ &  $\tau^{1/2}_{0\nu}
\langle m_\nu \rangle^2$
 [yr eV$^2$]  \\ \hline
$^{154}$Sm  &  4.898 10$^{-15}$ & 2.384  & 9.38 10$^{24}$  \\
$^{160}$Gd  &  1.480 10$^{-14}$ & 0.919  & 2.09 10$^{25}$  \\
$^{170}$Er  &  1.673 10$^{-15}$ & 0.731  & 2.92 10$^{26}$\\
$^{176}$Yb  &  6.817 10$^{-15}$ & 0.772  & 6.43 10$^{25}$\\
$^{232}$Th  &  3.160 10$^{-14}$ & 1.232  & 5.44 10$^{24}$ \\
$^{244}$Pu  &  1.463 10$^{-13}$ & 1.171  & 1.30 10$^{24}$ \\
\end{tabular}
\end{center}
\label{0nu}
\caption{The  phase-space integrals,  matrix elements and
predicted half-lives for the $0\nu$ double beta decay}
\end{table}

As a consequence of the explicit inclusion of deformation in the
present model, the \bz ~ half-lives are larger than those reported
in \cite{Sta90}. In \gd ~ the \bz -decay half-life is at least
three orders of magnitude larger than the \bt -decay half-life. It
implies that the background suppression due to a large \bt
~half-life would be effective, although not as noticeably as was
optimistically envisioned in \cite{Dan00}. In any case, the
results presented strongly suggest that the planned experiments
using GSO crystals \cite{Dan00} would be able to detect the \bt
~decay of \gd , and to establish competitive limits to the \bz
~decay.

The present results consider only the dominant pseudo SU(3) irrep
for each configuration. We have learned from realistic
calculations, where the single particle term and pairing
interactions induce the mixing of different irreps, that the
leading irreps represent in most even-even heavy deformed nuclei
at least 60 \% of the total wave function \cite{Var00a}. The
inclusion of spin dependent terms in the Hamiltonian,  relevant to
the description of the Gamow-Teller resonance, is not expected to
strongly modify the ground state wave function of the even-even
initial and final nuclei. This dominance lead us to expect that
future calculations, which will take into account contributions
from various irreps, would slightly affect the present
predictions. Given the leading role play by the
quadrupole-quadrupole interaction in heavy deformed nucleus, we
are confident that the order of magnitude of the predicted \bb
~half-lives, when various irreps are included in the calculations,
will remain unchanged, as compared to the results reported above.

Work supported in part by CONACyT, and by a CONACyT-CONICET
agreement under the project {\em Algebraic methods in nuclear and
subnuclear physics.} O.Civitarese is a fellow of the CONICET,
Argentina.

\end{document}